\documentclass[letterpaper]{article}
\usepackage{aaai}
\usepackage{times}
\usepackage{helvet}
\usepackage{courier}
\usepackage{graphicx}
\usepackage{tabularx}
\usepackage{multirow}
\usepackage{booktabs}
\usepackage{subfig}
\usepackage{makecell}
\graphicspath{ {./images/} }

\begin{document}
%
\title{Check Mate: Prioritizing User Generated Multi-Media Content for Fact-Checking}
\author{Tarunima Prabhakar, Anushree Gupta, Kruttika Nadig, Denny George \\
Tattle Civic Technologies\\
tarunima@tattle.co.in, anushree.gupta93@gmail.com, kruttika17@gmail.com, denny@tattle.co.in}

\maketitle
\begin{abstract}

Volume of content and misinformation on social media is rapidly increasing. There is a need for systems that can support fact checkers by prioritizing content that needs to be fact checked. Prior research on prioritizing content for fact-checking has focused on news media articles, predominantly in English language. Increasingly, misinformation is found in user-generated content. In this paper we present a novel dataset that can be used to prioritize check-worthy posts from multi-media content in Hindi. It is unique in its 1) focus on user generated content, 2) language and 3) accommodation of multi-modality in social media posts. In addition, we also provide metadata for each post such as number of shares and likes of the post on ShareChat, a popular Indian social media platform, that allows for correlative analysis around virality and misinformation. The data is accessible on Zenodo (https://zenodo.org/record/4032629) under Creative Commons Attribution License (CC BY 4.0).

\end{abstract}

\noindent

\section{Introduction}

\begin{figure}[t]
\centering
\includegraphics[scale=0.2]{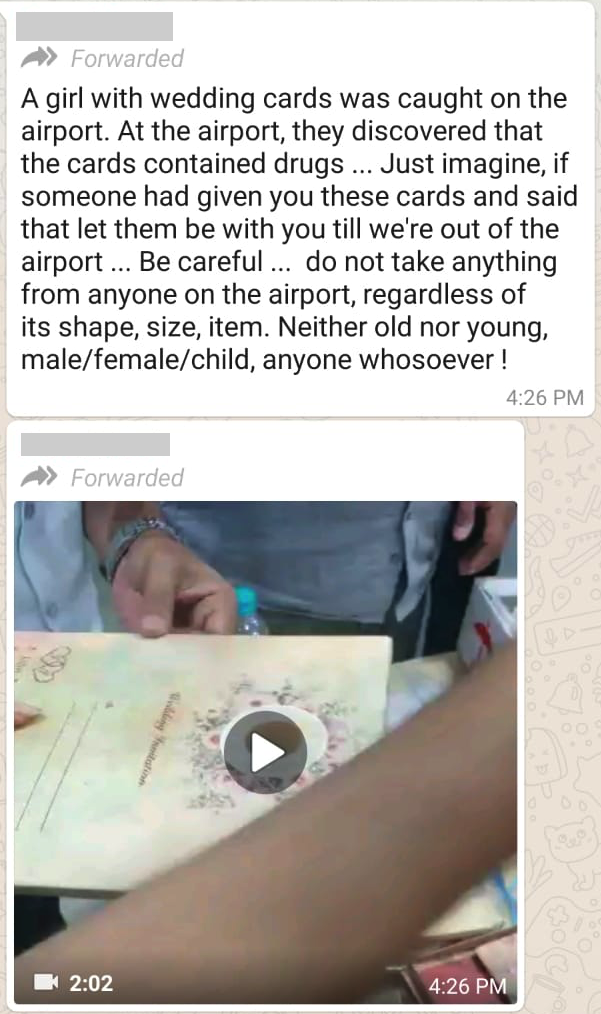}
\caption{False Context in Multi-Modal Content}
\label{fig:multimedia}
\end{figure}

Misinformation has emerged as an unfortunate by-product of growing social media usage. Over the last few years, rapid Internet penetration and mobile phone adoption in developing countries such as India has led to a rapid increase in social media users. Many such users use regional Indian languages as their primary language of communication. The resultant misinformation has had life threatening consequences \cite{whatsapp-vigilantes}. On chat apps such as WhatsApp, user-generated multi-media content has emerged as an important category of misinformation. Figure \ref{fig:multimedia} shows an example of a heavily circulated post on WhatsApp, where a video is recirculated with a misleading text message. The video and text together constitute a post that needs to be verified. 

Misinformation is contextual- the content and form of misinformation varies by topic of misinformation, platform of circulation, as well as the demographic targeted with that misinformation \cite{fake-news-data-mining}. The last few years have seen efforts towards characterizing and detecting `fake news' on social media. There is a consistent need for datasets emerging from diverse contexts to better understand the phenomenon. 

\begin{figure}[t]
\centering
\includegraphics[scale=0.2]{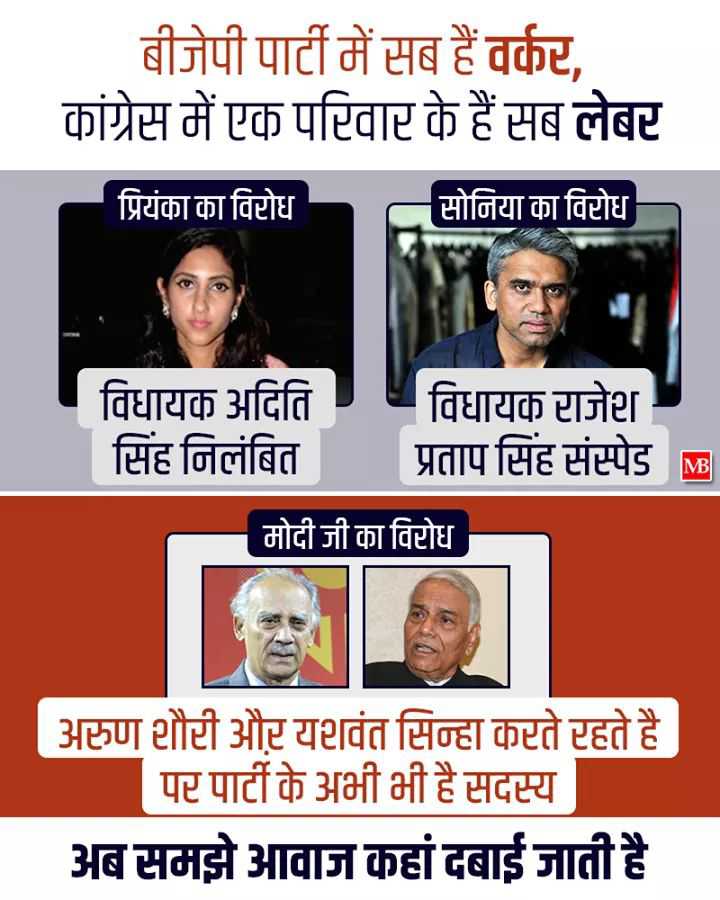}
\caption{Text Embedded in Image}
\label{fig:multimedia_2}
\end{figure}

Fact-checking has emerged as an important defense mechanism against misinformation. Social media platforms rely on third party fact checkers to flag content on them. Fact checking groups also run tiplines\footnote{https://techcrunch.com/2019/04/02/whatsapp-adds-a-tip-line-for-checking-fakes-in-india-ahead-of-elections/} on WhatsApp through which they can source content that requires fact checking. The amount of content received by fact checkers however, significantly exceeds journalistic resources. Automation can support fact checkers by flagging content that is of higher priority for fact-checking. Extracting claims and assessing check-worthiness of posts is one way to help prioritise specific content for fact-checking \cite{check-worthy-claims}. In case of news articles and speeches, these claims appear as factual statements made by writers or speakers. User generated content, however, is often multi-modal. Videos and images are supported with additional text or media to give false context, make false connections or create misleading content \cite{first-draft}. In Figure \ref{fig:multimedia}, the text message provides false context to an authentic video. In Figure \ref{fig:multimedia_2}, names and actions are attributed to people identified in images. In this specific case, the image of the individual in the top right is of a fashion designer and not of a politician, as claimed in the post.

Previous work has predominantly focused on claims detection for single media sources such as tweet texts, speeches and news articles \cite{check-worthy-claims,claim-detection-twitter}. Additionally, most research has focused on English language content. The rise in multi-modal user generated content makes it imperative to consider claims in multi-media posts. Furthermore, with regards to prioritizing multi-media content for fact checking, there is a need to deliberate on what is check-worthy in multi-media content. For example, selfie video claim lesser journalistic integrity than televised news coverage and might be of lower priority for fact checking.

In this paper, we present a unique dataset of multi-modal content in Hindi that can be used to train models for identifying check-worthy multi-modal content in the language. It is unique in its 1) focus on user generated content, 2) language and 3) claims detection in multi-modal content. In addition to the annotations, we also provide metadata for each post- such as time-stamp of creation; number of shares and likes; and the tags that accompany them on a popular Indian social media platform, ShareChat. Table \ref{datasummary} summarizes the different media types in the dataset. Table \ref{fieldssummary} provides an overview of the fields in the dataset. In the subsequent section, we provide background on data collection and discuss related work. We then describe the process of annotation, the results from the annotations; and possible use cases and limitations. 

\begin{table*}
\centering
\begin{tabularx}{\textwidth} { 
  | >{\centering\arraybackslash}X 
  | >{\centering\arraybackslash}X 
 |} 
\hline
  & Count \\[1ex]
\hline
 
Total Number of Posts & 2200 \\
\hline

Annotated by Single Person & 1800 \\
\hline

Annotated by Two People & 4000 \\
\hline

Posts with primary media type: Image & 1354 \\
\hline

Posts with primary media type: Video & 659 \\
\hline

Posts with primary media type: Text & 187 \\
\hline

\end{tabularx}
\caption{Dataset Summary}
\label{datasummary}
\end{table*}

\begin{table*}
\centering
\begin{tabularx}{\textwidth} { 
  | >{\centering\arraybackslash}X 
  | >{\centering\arraybackslash}X 
 |} 
\hline
Field Category & Field Names \\[1ex]
\hline

Fields Describing Post (from platform) & bucket name, media type, caption, tag name, tag translation, text\\
\hline

Annotated Fields & verifiable claim, contains video, contains image, visible source, contains relevant meme\\
\hline

Metadata Fields (from platform) & number of external shares, likes, estimated timestamp of creation\\
\hline

Other Fields  & Unique Post ID, Annotator Label\\
\hline

\end{tabularx}
\caption{Dataset Fields Overview}
\label{fieldssummary}
\end{table*}

\section{Related Work}

\subsection{Background on Sharechat}
The posts in this dataset were collected from an Indian social media platform called ShareChat. The platform has over 60 million monthly active users and is available in fourteen regional Indian languages \cite{mint_sharechat_2020}. It provides users the option to share other users' content on their profiles, as well as share content to other apps such as WhatsApp. \cite{sharechat_2020} provide a detailed summary of linguistic and temporal patterns observed in the content on the platform. For this dataset, we focus specifically on content in Hindi. With 500 million speakers, Hindi is the most commonly spoken Indian language \cite{census}.

\subsection{Characterizing Check-Worthiness}
Claim detection is a rich area of research \cite{benchmark-claim-detection,context-dependent-claim}.
Given their importance to the American electoral process, US presidential debates have received significant attention for  claim detection \cite{checkthat,tathya-2017,check-worthy-claims}. Some research has focused on claims in user generated content on Twitter. This work however is still focused on textual claims on the platform \cite{claim-detection-twitter,twitter-claims}.

The work of \cite{zlatkova-etal-2019-fact} is noteworthy for its focus on textual claims about images. In the field of misinformation detection, multi-media content has received more attention in the last few years. \cite{nakamura-etal-2020-fakeddit} provide a dataset of one million multi-modal posts from Reddit, classified into different types of misinformation. \cite{whatsapp-images-election} provide a dataset of Brazilian and Indian election related images circulated on WhatsApp, that were found to be misinformation. \cite{sandy-images} analyze fake images on Twitter during Hurricane Sandy. 

More broadly, literature describing and classifying misinformation highlights features that could help in prioritization of multi-media user-generated content for fact checking. \cite{fake-news-data-mining} focus specifically on fake news articles which contain two major components- publisher and content. They classify news features using four attributes: source, headline, main text and image/video contained in the article. In their characterization of fake news, the image/video is considered as a supporting visual cue to the main news article. This is different from user generated content on chat apps, where the text can be supplementary context for an image or video. \cite{taxonomy-online-content} propose a eight category typology for identifying 'fake news' which ranges from citizen journalism to satire and commentary. They further identify content attributes such as factuality, message quality, evidence, source and intention to characterize these eight categories. 

Since a lot of misinformation research has focused on English content, we also highlight here work that focuses on non-English media. The dataset provided by \cite{whatsapp-images-election} has images with content in Portugese and Indian languages. \cite{language-independent-fake-news} look at language features across English, Spanish and Portugese for fake news detection. 

\section{Data Collection}
\begin{figure}
\centering
\includegraphics[scale=0.31]{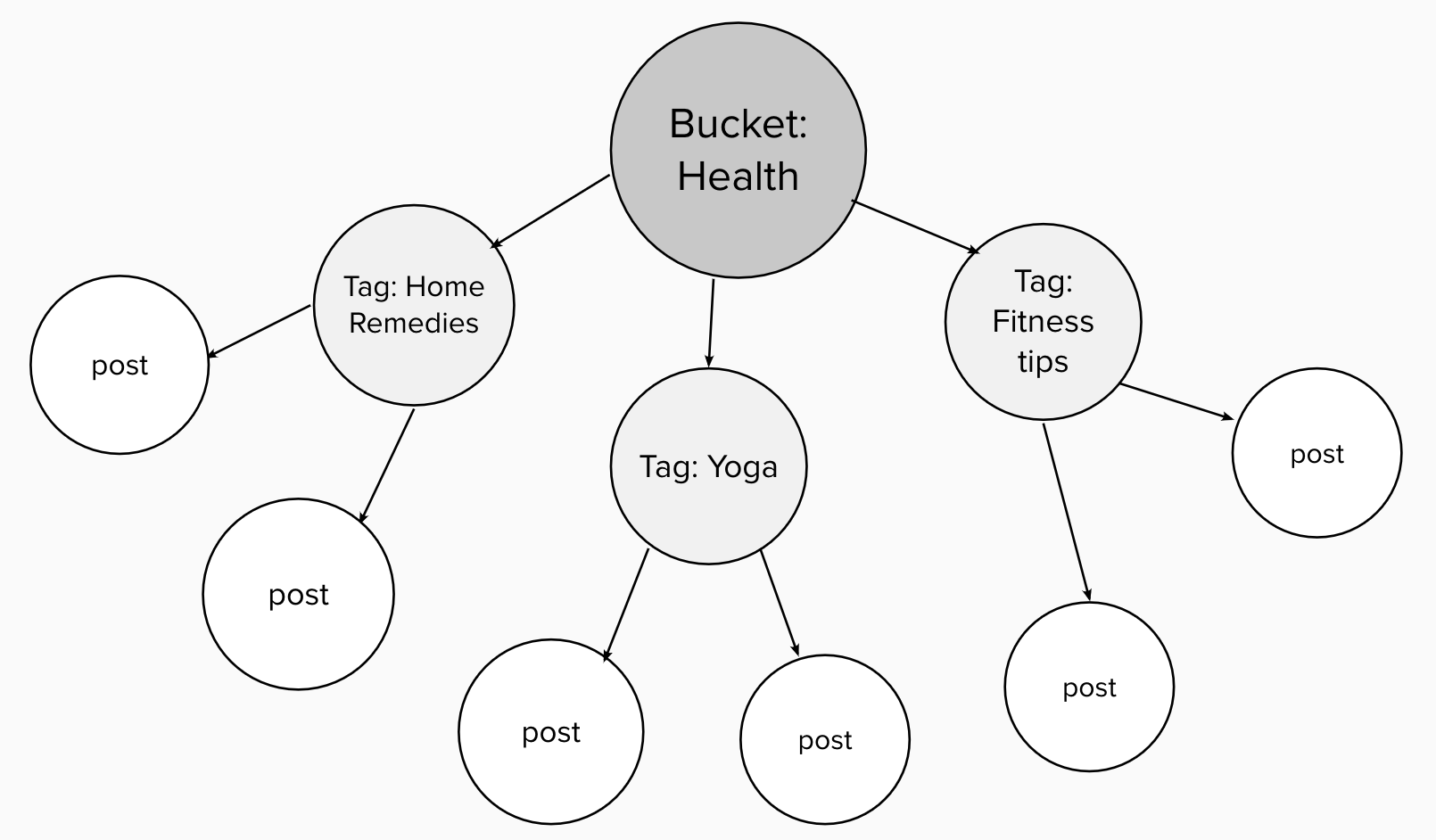}
\caption{ShareChat Content Tree Structure}
\label{fig:sharechat}
\end{figure}

To source user-generated multi-media content, we scraped content from ShareChat. On ShareChat, content is organized by `buckets' corresponding to topic areas. Each bucket has multiple tags. Figure \ref{fig:sharechat} shows the conceptual organization of content on the platform. We scraped content daily from the `Health', `Coronavirus' and `Politics and News' buckets in Hindi from March 21 2020 to August 4 2020. We focused on these buckets since politics and health have been observed to be the most common themes of misinformation in India \cite{bbc_duty_2018}. This resulted in over 200,000 unique posts. We randomly sampled from this bigger corpus to prepare the dataset presented here.
The platform provides a number of additional data points such as the number of shares and likes, that provide some social context for a post. The platform also assigns a `media type' to each post which could be text, image, video or link. Regardless of the media type, there is an option for users to provide captions for a post, which can contain large text descriptions. For our sampling, we ignore posts of media-type `link' since these redirect to other websites and are therefore not within the purview of our focus on user generated content on the platform. 

\section{Qualitative Analysis}
\subsection{Open Coding}
We use `Open coding' methodology to identify features that can help prioritize multi-media content for fact checking. Open Coding is an analytic qualitative data analysis method to systematically analyze data \cite{open-coding}.  In the first round, two researchers from the team loosely annotated and described 600 posts to familiarize themselves with the data. In the second phase, we described 300 posts in detail. We converged on sixteen non-mutually exclusive categories of multi-media posts that could be verified. These included posts that contained statistical claims (such as Covid fatality numbers); quotes attributed to noteworthy individuals; and newspaper clips and edited recordings from televised news reporting. 

We also recognized user generated videos claiming to debunk misinformation as a separate category of content- the period of data collection coincided with Covid-19 and some posts were warnings about misinformation on the pandemic.

From these sixteen categories we converged on the following broad attributes of check-worthy content:
 \begin{itemize}
     \item The implied/explicit source of the content: 
     Similar to \cite{fake-news-data-mining,taxonomy-online-content} we find the \emph{implied} source of the content to be an important attribute for prioritizing content for fact checking. We identified five common sources of data (see Table \ref{annotation}) that are ascribed differing levels of authenticity. Circulars and messages attributed to government sources for example are more likely to be believed than anonymous sources.
     
     \item The kind of verifiable claim:
     \cite{check-worthy-claims} specifically characterize sentences from US presidential debates into three categories- non-factual statement, unimportant factual statement and check-worthy factual statement. Since this dataset is multi-modal, we found it useful to further qualify check-worthy factual statements. We define three broad categories of check-worthy claims in multi-media content: 1. statistical/numerical claims. 2.  world events or places about noteworthy individuals. 3. Other factual claims such as home remedies, cures and nutritional advice. 
     
     \item For video content, the `intentionality' of creation, as inferred from the video:
     There are multiple ways to categorize user generated video content. We broadly divide the content into two groups- videos where the person in the video is performing for the camera, and videos where the recording is incidental to the event. Given the ability to provoke, we find it useful to highlight violent incidents portrayed in videos as a third category. Notably, we do not consider an image montage to be a video, even if the media type as declared by the platform is 'video'.
     
     \item Specifically for non digitally-manipulated images, the topic of images:
     While a lot of content on ShareChat is classified as `image', not all of it is unidentified. A lot of `Good Morning Wishes' memes use generic background images. In our framework, such images are not considered important for fact checking. We draw a distinction between digitally manipulated images and non-digitally manipulated images. 
     
     We realize that within non-digitally manipulated images, the topic of the image is an important feature that could help prioritize a content for fact checking. This however would require developing a content taxonomy of social media content. This is an important and complex research area and merits dedicated attention. We found this taxonomy to be beyond the ambit of this specific work. We loosely classified non-digitally manipulated images into those that contain human figures and those that don't. We found it useful to flag images with human figures in them, since the context ascribed to the photograph can be verified.
     
     \item Culturally Relevant Memes: In the absence of a relevant taxonomy, we defined a broad binary category for `relevant memes'. For the purposes of this annotation, relevant memes are those that reference security forces, religious pride, national pride, political party representatives, reverential posts about a profession, election campaigning or current national news. While a more detailed taxonomy is needed, this broad descriptor of culturally important themes, when combined with other attributes such as source and verifiable claims, can become important in highlighting a post that needs to be fact-checked.
 \end{itemize}

\begin{figure*}[h]
    \centering
    \subfloat[]{\includegraphics[scale=0.15]{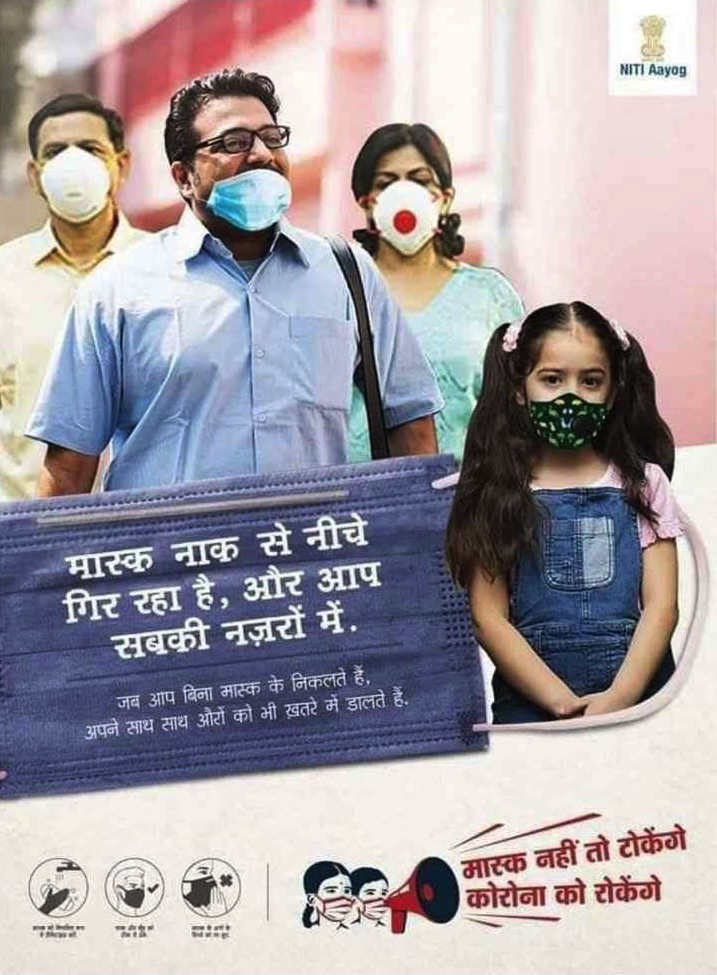}}
    \hspace{10mm}
    \subfloat[]{\includegraphics[scale=0.15]{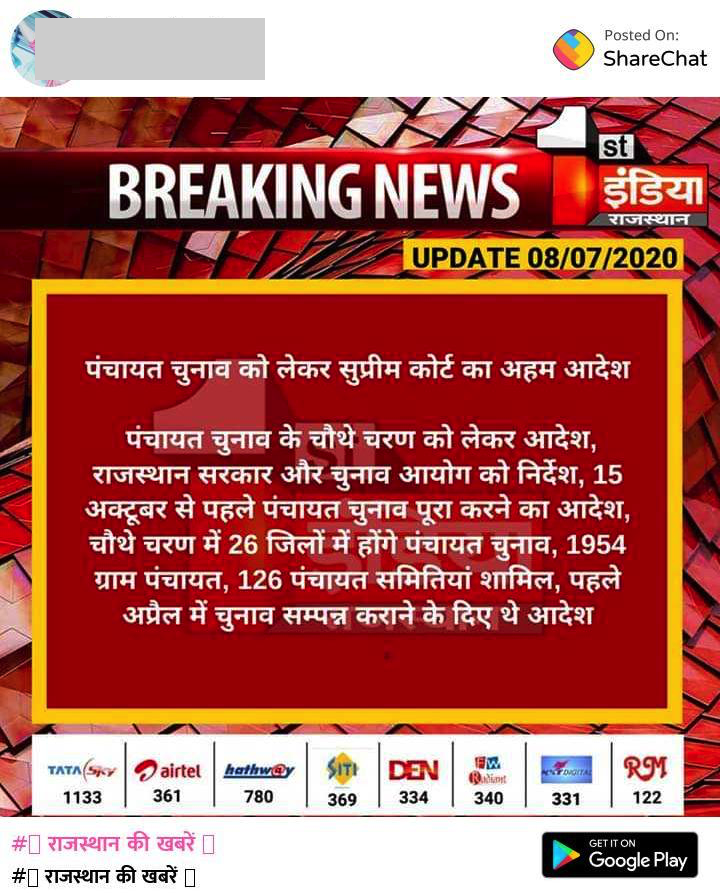}}
    \caption{(a) Post Attributed to Government Body (Niti Aayog) (b) A Screen Grab of TV News With Factual Claims}
    \label{fig:samples_1}
    
\end{figure*}

\begin{figure*}[h]
    \centering
    \subfloat[]{\includegraphics[scale=0.35]{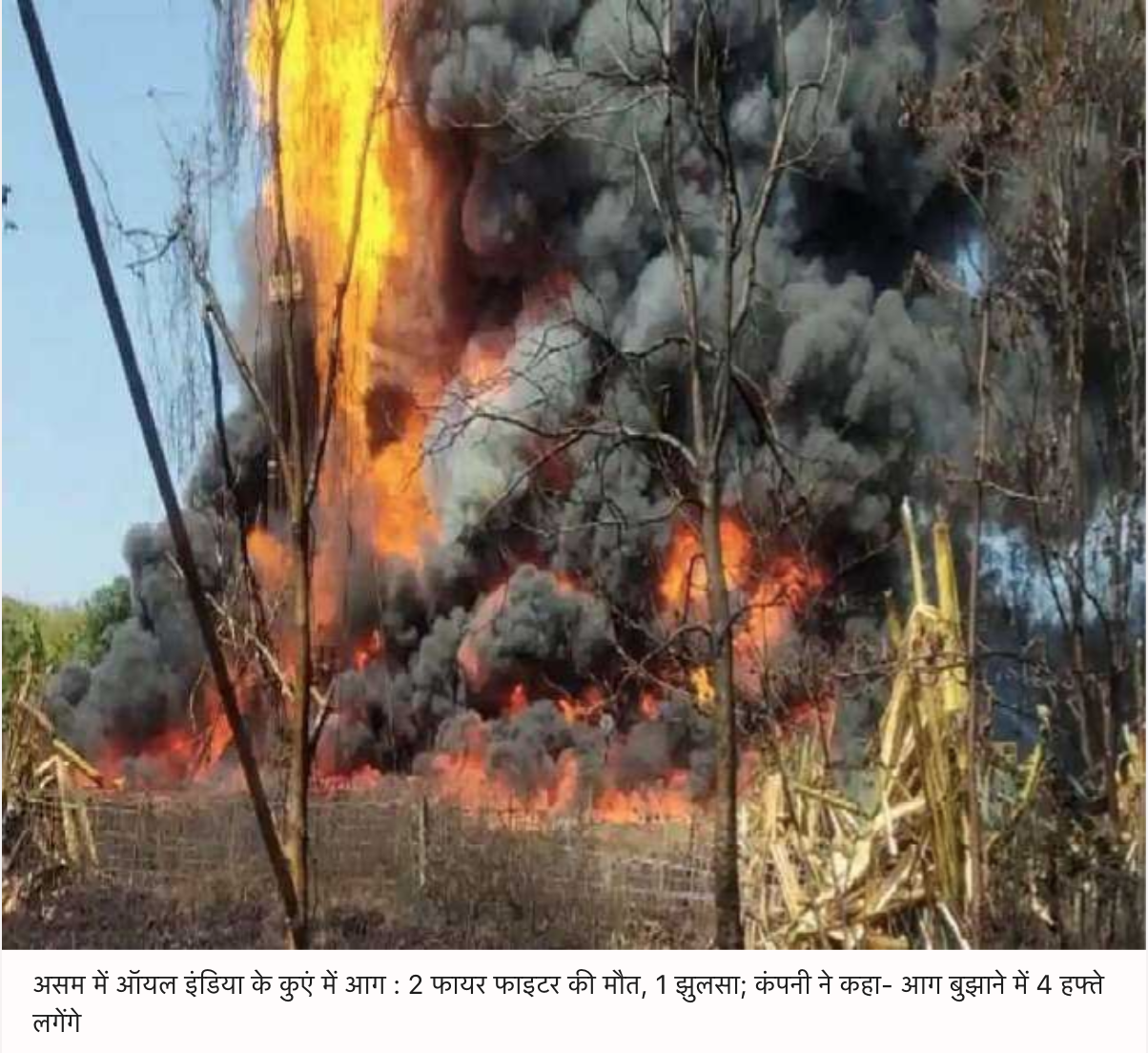}}
    \hspace{5mm}
    \subfloat[]{\includegraphics[scale=0.3]{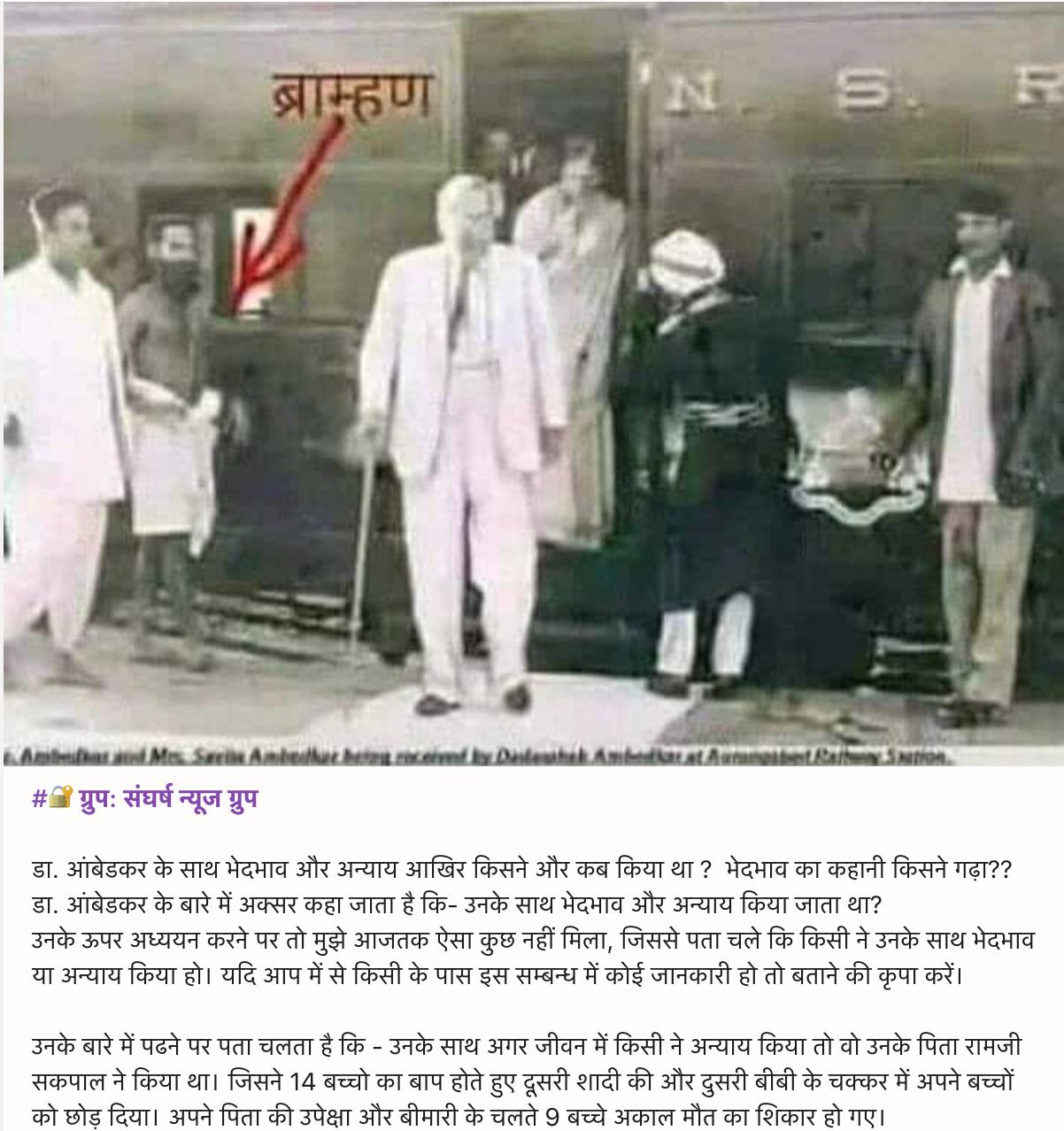}}
    \caption{(a) An Unidentified Image with Text Description About an Oil Well Accident (b) An Image with Text Description about Dr. Bhimrao Ambedkar}
    \label{fig:samples_2}
\end{figure*}

Table \ref{annotation} summarizes these five attributes. 

\begin{table*}[h!]
\centering
\begin{tabular}{|p{12em}|p{35em}|}
 \hline
 Attribute & Options\\ [1ex] 
 \hline 

 \multirow{6}{12em}{Implied or Explicit Source of Content(mutually-exclusive)} 
        &\parbox[c][1.5cm]{35em}{\textbf{Print/TV Media:} 
        Newspaper clips or screen grabs from print or TV media sources. Examples of TV media are BBC, CNN, Aaj Tak (Indian); examples of print media: The Times of India, Dainik Jagran, Rajasthan Times. See Figure \ref{fig:samples_1}.b} \\\cline{2-2}
        
        &\parbox[c][0.75cm]{35em}{\textbf{Digital Only News Outlets} such as Huffington Post, Editor ji, myUpchar.} \\\cline{2-2}
        
        &\parbox[c][1cm]{35em}{\textbf{Digital Content Providers} such as Buzzfeed Goodful, or Instagram and Facebook pages.}\\\cline{2-2}
        
        &\parbox[c][2cm]{35em}{\textbf{Government/Public Authority:} Government logos or stamps visible in the post or content attributed to spokespersons, officials, civil servants/bureaucrats, other public office holders.
        OR  Multilateral institutions and international public bodies such as WHO, ILO and World Bank. See Figure \ref{fig:samples_1}.a} \\[10pt]\cline{2-2}
        
        &\parbox[c][0.5cm]{35em}{Anonymous/Social Media Users. See Figure \ref{fig:samples_2}.a} \\\cline{2-2}
        
        &\parbox[c][0.5cm]{35em}{Other} \\ 

\hline

\multirow{3}{12em}{Type of Factual Claim (not mutually-exclusive)} 
    &\parbox[c][0.5cm]{35em}{Statistical or numerical claim. See Figure \ref{fig:samples_1}.a}\\\cline{2-2}
    &\parbox[c][0.5cm]{35em}{Descriptions of real world places, events or a noteworthy individual(s). See Figure \ref{fig:samples_2}.b} \\\cline{2-2}
    &\parbox[c][0.5cm]{35em}{Other factual claims} \\ 
    
\hline
    
\multirow{3}{12em}{ Intentionality Portrayed in the Video (mutually-exclusive)} 
    &\parbox[c][0.5cm]{35em}{Person is Performing for Camera} \\\cline{2-2}
    &\parbox[c][0.5cm]{35em}{Contains Violent Incident} \\\cline{2-2}
    &\parbox[c][0.5cm]{35em}{Other}\\ 

\hline

\multirow{2}{12em}{Non Manipulated Images} 
    &\parbox[c][0.5cm]{35em}{Contains Humans. See Figure \ref{fig:samples_2}.b and Figure \ref{fig:samples_1}.1 } \\\cline{2-2} 
    &\parbox[c][0.5cm]{35em}{Does not Contain Humans. See Figure \ref{fig:samples_2}.a} \\\hline 
    
    Contains Relevant Memes? & \parbox[c][0.5cm]{35em}{Yes/No}  \\ 
\hline
\end{tabular}
\caption{Annotation Guideline}
\label{annotation}
\end{table*}

\subsection{Closed Coding}
Based on the above mentioned guidelines, the two researchers independently labeled posts in batches of fifty to refine these definitions. 
The framework developed was expanded into an annotation guideline with descriptions of each of the attributes to be labeled as well as sample annotations for thirteen posts. The annotation guide is provided with the dataset.
The annotation guideline was opened to two additional members of the Tattle team, who were not a part of the open coding stage. By virtue of their work, each member is familiar with the misinformation challenge in India, but not equally familiar with the nature of the content circulated. The four person annotation team was then split into two pairs. To estimate agreement across annotators, each person in the pair annotated 200 hundred posts common with the other member of the pair.
Table \ref{agreement} lists the agreement scores for the two pairs of annotators across the different attributes. The agreement for video type and image type attribute was substantial (greater than 0.6) for both pairs. Annotators strongly agreed on a post containing a verifiable claim as well. This can be used to train models for claim detection on multi-media posts. There was far weaker agreement on the specific kind of factual claim, when a post did contain one. Finally, on the implied or explicit source of content, one pair had moderate agreement while the second had weak agreement.

\begin{table*}[t]
\centering
\begin{tabularx}{\textwidth} { 
  | >{\centering\arraybackslash}X 
  | >{\centering\arraybackslash}X 
  | >{\centering\arraybackslash}X 
  | >{\centering\arraybackslash}X 
  | >{\centering\arraybackslash}X 
  | >{\centering\arraybackslash}X 
  | >{\centering\arraybackslash}X 
  | >{\centering\arraybackslash}X 
  | >{\centering\arraybackslash}X |} 
\hline
 & Video Type & Image Type & Source & No Factual Claim & Statistical Claim & Real Event, Place, Person & Other Factual Claim & Relevant Memes\\
\hline
 Pair 1 & 0.74 & 0.68 & 0.5 & 0.76 & 0.65 & 0.30 & 0.54 & 0.70\\
\hline
 Pair 2 & 0.85 & 0.68 & 0.35 & 0.83 & 0.33 & 0.31 & 0.50 & 0.61\\
\hline
\end{tabularx}
\caption{ Agreement}
\label{agreement}
\end{table*}

\section{Annotation Procedure}
\begin{figure*}[t]
    \centering
    \includegraphics[scale=0.3]{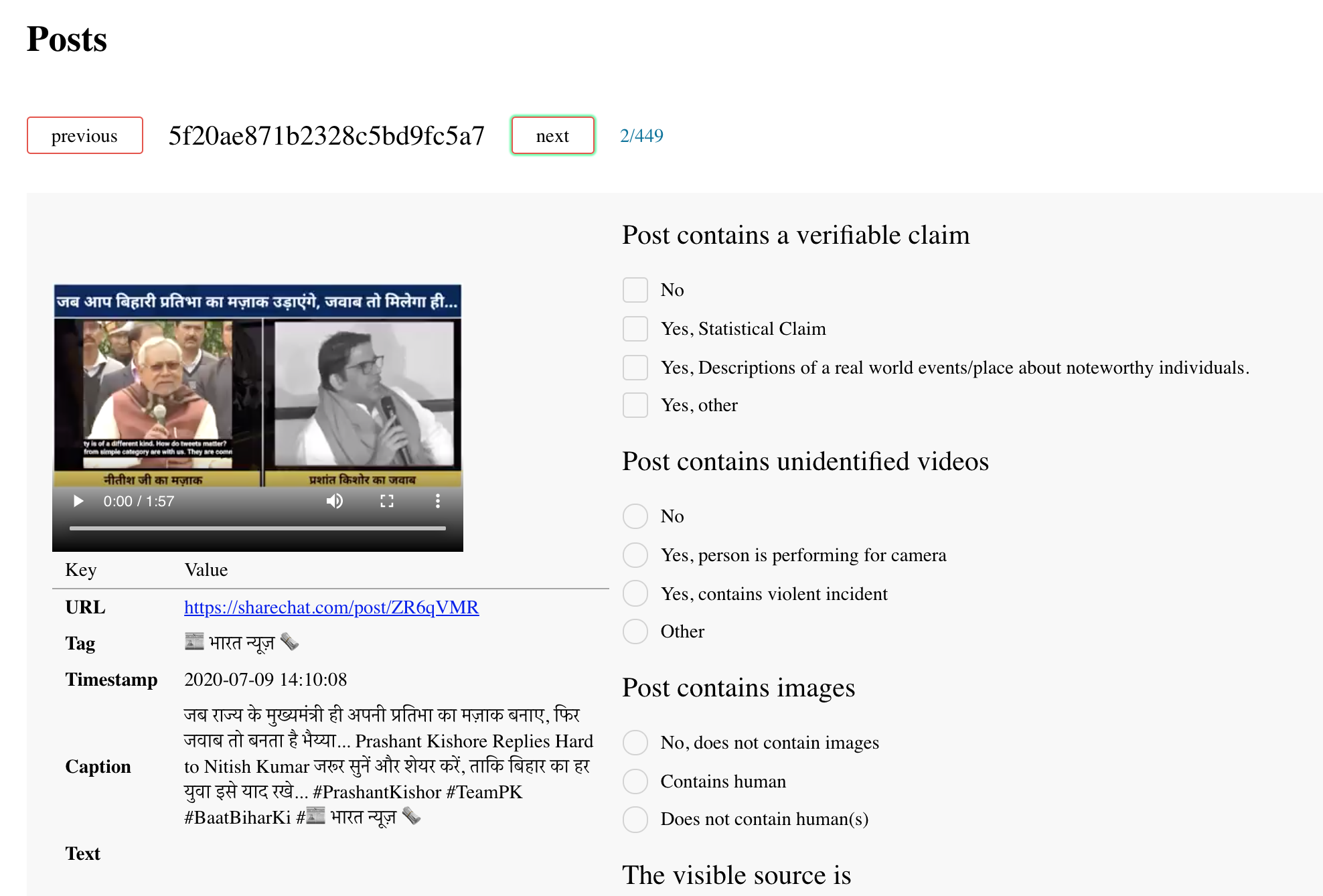}
    \caption{Annotation Interface}
    \label{fig:annotationUI}
\end{figure*}

After observing moderate to high agreement across four fields, each annotator proceeded to annotate 450 additional posts. This resulted in a total of 1800 posts uniquely annotated by one of the four annotators. We developed an in-house user interface for annotating specific posts\footnote{http://annotation.test.tattle.co.in.s3-website.ap-south-1.amazonaws.com/}. The interface is a Gatsby App built using React and Grommet UI library. Through the interface, annotators could load post related data from a JSON file and preview the media and metadata associated with it. Annotators could navigate from one post to another using `previous' and `next'. On clicking `next' annotated labels gets saved in the same JSON file. (See figure \ref{fig:annotationUI})

Attributes such as video style, image content and source that had mutually exclusive options were enforced as single option questions in the interface. The attribute about the kind of factual claim was listed as a check-box, since a post can contain multiple kinds of factual claims.

After all the annotations were completed, the JSON files were converted to CSV files for greater legibility and ease of sharing.

\section{Dataset Description} \label{data sample}
\subsection{Data Sample} 
In this dataset we share 400 posts that were annotated by two people each in the closed coding phase, as well as 1800 posts annotated by four separate annotators. In total, there are 2200 annotated posts shared in this dataset.

Before expanding on the annotations, we'll describe the data fields scraped from ShareChat. The metadata about a post scraped from the platform includes:
\begin{itemize}
    \item timestamp: time at which the post was created on the platform
    \item media type: can be image, video or text
    \item text: text content if media type of post is text.
    \item caption: text caption to images and videos. 
    \item external shares: number of times the post was shared from the platform to an external app such as WhatsApp
    \item number of likes: number of times the post was liked on ShareChat
    \item tags: the tags attached to the post by the user. See figure \ref{fig:sharechat}
    \item bucket name: the bucket under which this post was classified by the platform. See figure \ref{fig:sharechat}
\end{itemize}

 Regardless of the media type attributed by the platform, users can add additional text for images and videos which is captured in the `caption' field for each post. 

\begin{figure*}[h]
    \centering
    \subfloat[]{\includegraphics[scale=0.55]{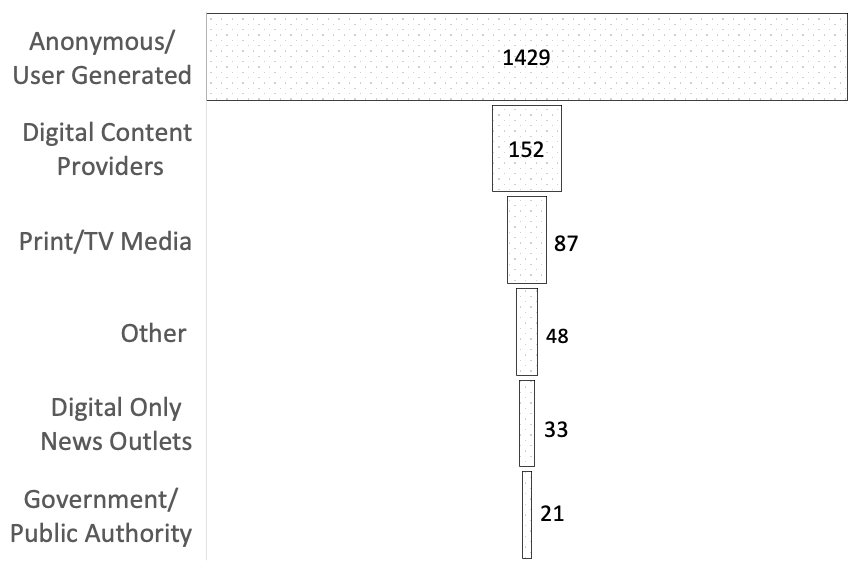}}
    \hspace{5mm}
     \subfloat[]{\includegraphics[scale=0.55]{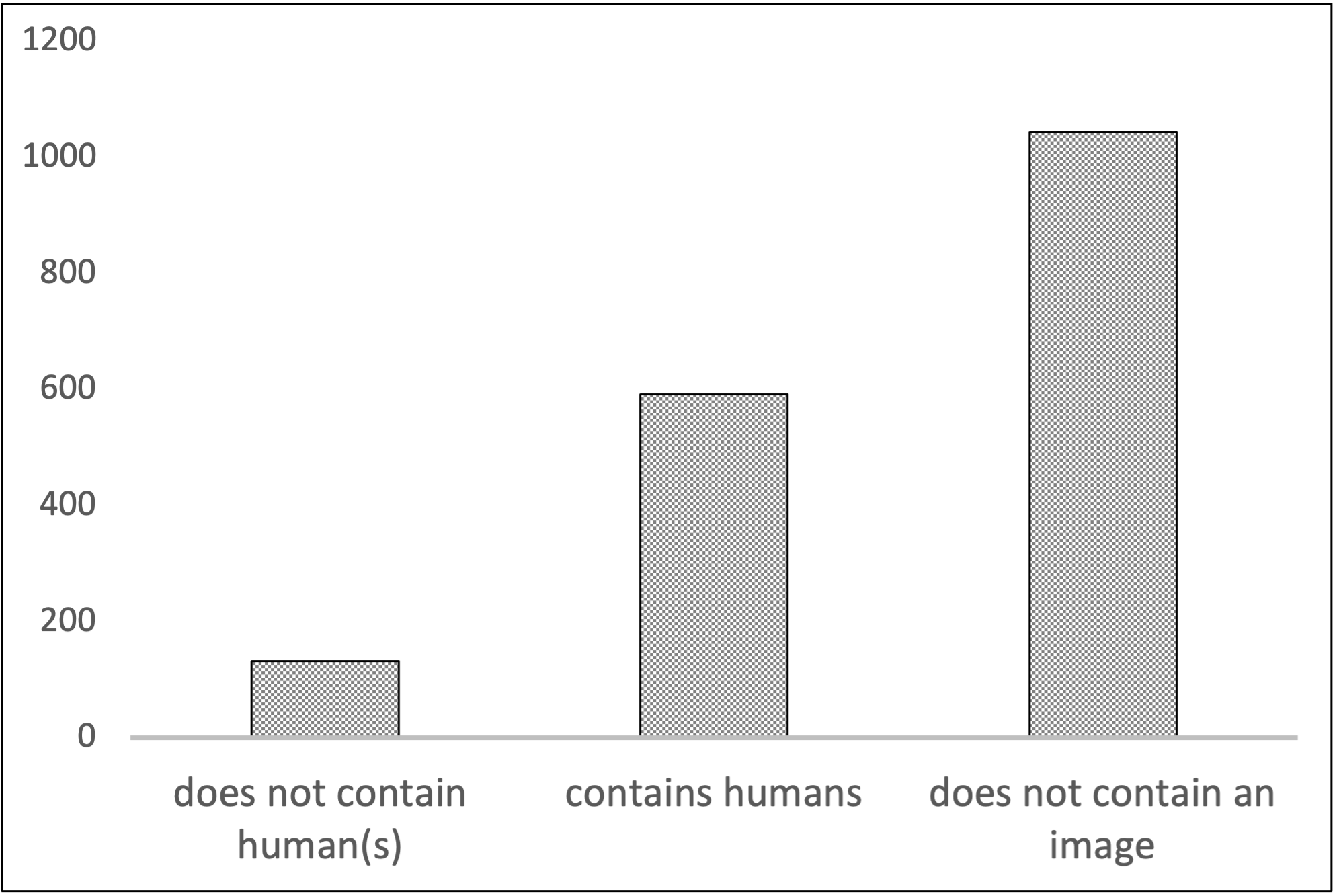}}
    \caption{(a) Visible Source (b) Type of Non-Manipulated Images}
    \label{fig:graphs_1}
\end{figure*}

\begin{figure*}[h]
    \centering
    \subfloat[]{\includegraphics[scale=0.55]{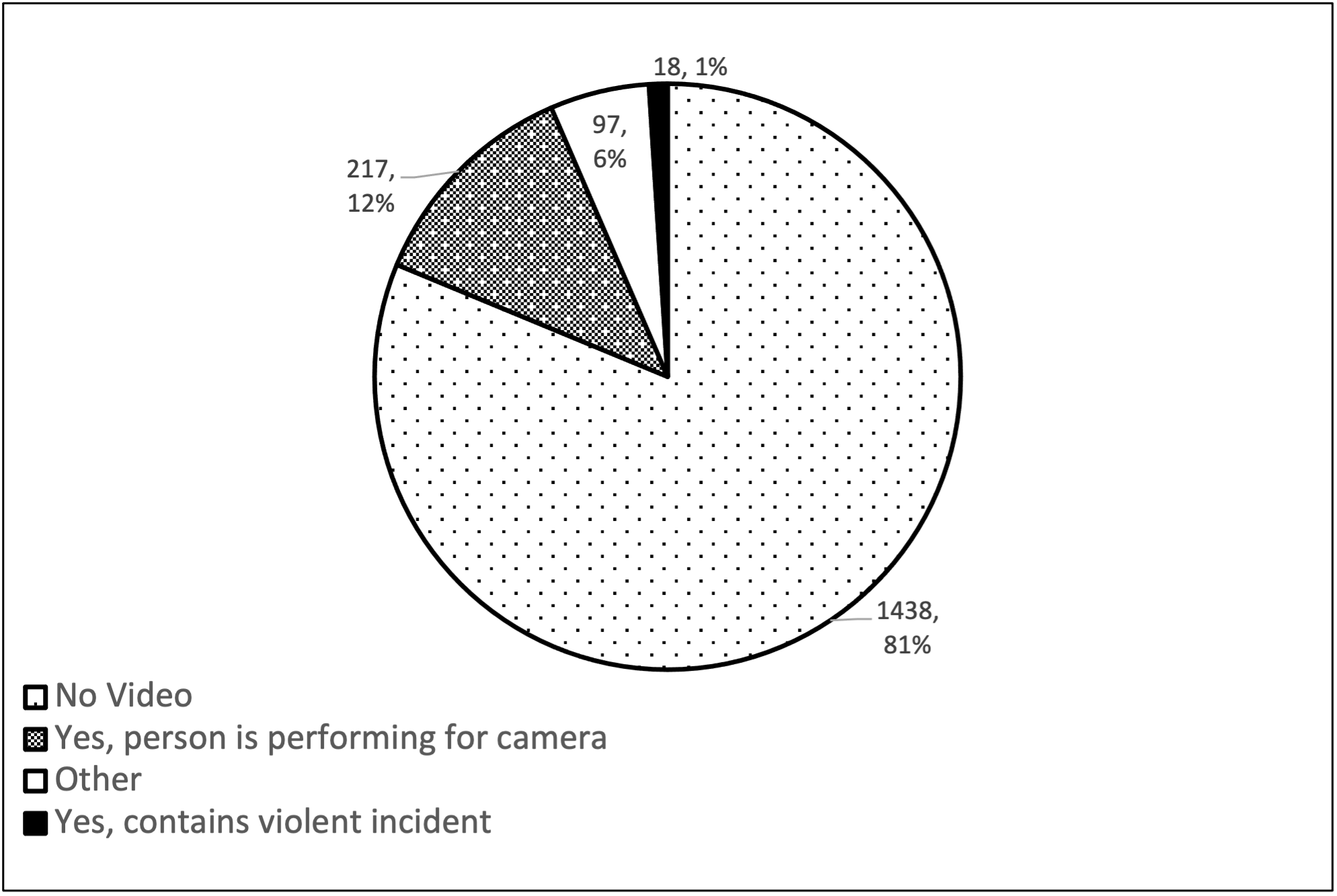}}
    \hspace{5mm}
    \subfloat[]{\includegraphics[scale=0.55]{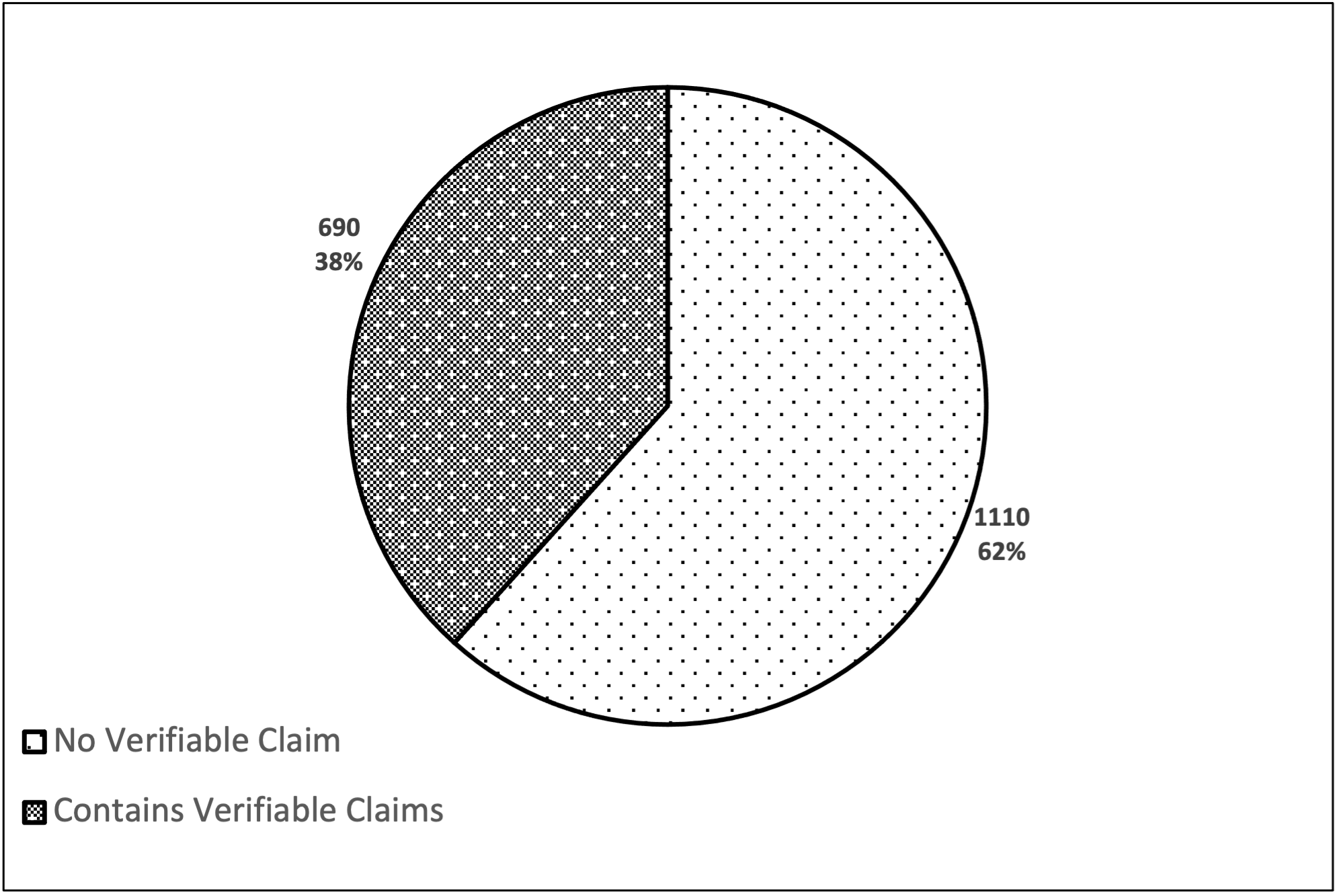}}
    \caption{ (a) Type of Video  (b) Proportion of Posts Containing Verifiable Claims}
    \label{fig:graphs_2}
\end{figure*}

\vspace{4mm}
\subsection{Annotated Data}
The fields developed in the closed coding stage were annotated as following:
\begin{itemize}
    \item verifiable claim: a list of kinds of verifiable claims contained in the post. If no claim was found, the post is labeled as `No'. The claims can be those listed in Table \ref{annotation}. Since a post can have more than one kind of claim, the annotated field is a list of claims. 
    \item contains video: if the post contains a video, it can be marked as `Person Performing for Camera' or `Contains Violent Incident'. A video that doesn't fall in these two categories is marked as other. A post that does not contain a video is marked as `No'.
    \item contains image: A post is said to contain an image if the image is \textbf{not} digitally manipulated. Within an image, a post can be marked as `containing human figures' or `not containing human figures'. A post that does not contain an image is marked as `No'.
    \item visible source: Marked as one of the sources listed in Table \ref{annotation}
    \item contains relevant meme: a binary category which is marked yes if the post is a text or image (not video) based meme about a culturally relevant topic.

\end{itemize}

Figure \ref{fig:graphs_2} summarizes the attributes of the annotated posts (n=1800) that had high agreement in closed coding stage. Even though a majority of posts are labeled as images by the platform, only forty percent of the posts have non-manipulated images that can be fact checked (see Figure \ref{fig:graphs_1}(b)). In this sample of data, for a majority of posts, the visible source of the post was the platform user (see Figure \ref{fig:graphs_1}(a)). It must be clarified that the user could have picked up their content from other established sources such as online websites and newspapers. Those sources however were not visible in the post. Part of the challenge of fact-checking user generated content is identifying sources from which content emerges. The high prevalence of user generated content without a recognized source, emphasizes the need for systems that prioritize this content for fact checking.  

\section{Dataset Structure}

The Tattle CheckMate Database consists of two folders. The first folder contains annotations from the closed coding phase. The second folder contains annotations of 1800 posts done by single annotators. The dataset also contains a README file that describes the folder structure and fields in the datasets.

\section{Ethical Considerations}
\subsection{Data Collection and Sharing}
While content creators need to be logged into the platform to upload posts, anyone with access to the post URL can view it. We thus treat the posts as content in the public domain that can be scraped and researched. We recognize that this content may contain images and videos of platform users and that opening the data can introduce some risk of identification of people in these posts. We have however mitigated this risk by removing the original source URL and sharing only the post. We also recognize that the violent imagery in some of the posts could be disturbing for users of this data. While labels such as those for violent videos could serve as an alert or trigger warning for some posts, we recognize that this is not comprehensive. Disturbing imagery is however unavoidable in the problem space that this dataset aims to intervene in. It is, in fact, one aspect of social media use that we hope this dataset can inform.

\subsection{FAIR Requirements}
This dataset adheres to FAIR principles (Findable, Accessible, Interoperable and Re-usable).
The dataset is Findable and Accessible through Zenodo on https://zenodo.org/record/4032629 and licensed under Creative Commons Attribution License (CC BY 4.0). The data is shared as a set of CSV files with the referenced media items in the folder. Thus the data is reusable and inter-operable.

\section{Conclusion}
To the best of our knowledge, this dataset is a first of its kind in focusing on claims detection and prioritization of multi-media content for fact checking. This dataset provides a starting framework for features in different mediums that can help in flagging content. Future work can expand on this framework for more granular features across the different modalities.
This dataset is also novel in its cultural and linguistic context and can be used for a range of open research problems:

\subsection{Training Models to Prioritize Multi-modal Content for Fact-Checking/Verification}
This dataset can be used for building models that can highlight check-worthy multi-modal content, especially in Hindi. This data can also supplement other datasets on misinformation.  \cite{whatsapp-images-election} summarize datasets that are focused on the automation of fake news detection.  

\subsection{Understanding Content Virality}
This dataset provides content features as well as some social context such as the number of shares, likes and tags. It can be used to better understand the relationship between the content and style of post and its popularity.

This work provides a framework for working with multi-modal content, but it also highlights a number of open research questions such as the need to build a taxonomy of image based content, and a more detailed typology of verifiable claims that are relevant for prioritizing of multi-modal content for fact checking. Future research can expand on this work to support such questions with other datasets.

\section{Acknowledgements}
We thank Dr. Sandeep Avula and Dr. Swair Shah for their continuous feedback in creation of this dataset. This work was partially supported by the `The Ethics and Governance of the Artificial Intelligence Fund' of the AI Ethics Initiative.

\bibliographystyle{aaai}\bibliography{bibfile}

\end{document}